\documentclass[preprint,superscriptaddress]{revtex4-1}
\usepackage{graphicx}
\usepackage{bm}
\usepackage{amsmath}
\usepackage{amsfonts}
\usepackage{amssymb}
\usepackage[utf8]{inputenc}
\newcommand {\be} {\begin{equation}}
\newcommand {\ba} {\begin{eqnarray}}
\newcommand {\ee} {\end{equation}}
\newcommand {\ea} {\end{eqnarray}}
\begin{document}

\title{Elastic pion-proton and pion-pion scattering via the holographic Pomeron and Reggeon exchange}

%---------------------------------------------------------------------------------------------------
\author{Zhibo Liu}
\email{202107020021014@ctgu.edu.cn}
\affiliation{College of Science, China Three Gorges University, Yichang 443002, People's Republic of China}
%---------------------------------------------------------------------------------------------------
\author{Akira Watanabe}
\email{watanabe@ctgu.edu.cn (Corresponding author)}
\affiliation{College of Science, China Three Gorges University, Yichang 443002, People's Republic of China}
\affiliation{Center for Astronomy and Space Sciences, China Three Gorges University, Yichang 443002, People's Republic of China}
%---------------------------------------------------------------------------------------------------

\date{\today}

%%%%%%%%%%%%%%%%%%%%%%%%%%%%%%
\begin{abstract}
The elastic pion-proton and pion-pion scattering are studied in a holographic QCD model, focusing on the Regge regime.
Taking into account the Pomeron and Reggeon exchange, which are described by the Reggeized $2^{++}$ glueball and vector meson propagator respectively, the total and differential cross sections are calculated.
The adjustable parameters involved in the model are determined with the experimental data of the pion-proton total cross sections.
The differential cross sections can be predicted without any additional parameters, and it is shown that our predictions are consistent with the data.
The energy dependence of the Pomeron and Reggeon contribution is also discussed.
\end{abstract}
%%%%%%%%%%%%%%%%%%%%%%%%%%%%%%

\maketitle

%%%%%%%%%%%%%%%%%%%%%%%%%%%%%%%%%%%%%%%%%%%%%%%%%%%%%%%%%%%%%%%%%%%%%%%%%%%%%%%
\section{Introduction}
\label{sec:introduction}
%%%%%%%%%%%%%%%%%%%%%%%%%%%%%%%%%%%%%%%%%%%%%%%%%%%%%%%%%%%%%%%%%%%%%%%%%%%%%%%
Studying elastic hadron-hadron scattering at high energies is one of the important research topics in high energy physics, since the cross sections include the information for the internal structure of the involved hadrons.
However, since those cross sections in the quantum chromodynamics (QCD) processes are basically nonperturbative physical quantities, it is difficult to perform the analysis by the direct use of QCD.
Although in some quite limited kinematic region, such as the high energy limit, the perturbative QCD is available and important results have been obtained~\cite{Brodsky:1973kr, Matveev:1973ra, Lepage:1980fj}, for most kinematic region effective approaches are required to calculate the cross sections.
It is particularly difficult to perform the theoretical analysis on the high energy forward scattering, since the underlying partonic dynamics is highly nonperturbative.
In this study we focus on the elastic pion-proton ($\pi p$) and pion-pion ($\pi\pi$) scattering in the forward region, and investigate the total and differential cross sections in the framework of holographic QCD, which is one of the effective approaches for QCD.
The pion is a special particle, since it is identified as the lightest Nambu-Goldstone boson.
It is important to understand the structure of the pion to deepen our understanding of the strong interaction, and the pion-nucleon and pion-pion scattering have been studied by various researchers so far~\cite{Pelaez:2003ky, Bzdak:2007qq, Petrov:2009wr, Halzen:2011xc, Caprini:2011ky, Mathieu:2015gxa, Pelaez:2015qba, Ryutin:2016hyi, Khoze:2017bgh}.

Historically, it is known that the Regge theory can give reasonable descriptions for various high energy forward scattering processes in the Regge regime, in which the condition, $s \gg t$ ($s$ and $t$ are the Mandelstam variables), is satisfied.
It is based on the analysis with the complex angular momentum, and this theory has been successfully applied to the hadron-hadron scattering~\cite{COLLINS1971103, Collins:1977jy}.
In the Regge theory, the scattering amplitudes can be obtained by considering the Pomeron and Reggeon exchange, which can be interpreted as the multi-gluon and meson exchange, respectively.
In the medium energy region, the both contributions are substantial, and the contribution of the Pomeron exchange becomes dominant in the high energy region.

The description with the Pomeron and Reggeon can also be realized in the framework of holographic QCD~\cite{Kruczenski:2003be, Son:2003et, Kruczenski:2003uq, Sakai:2004cn, Erlich:2005qh, Sakai:2005yt, DaRold:2005mxj, Brodsky:2014yha}, which is constructed based on the anti-de Sitter/conformal field theory (AdS/CFT) correspondence~\cite{Maldacena:1997re, Gubser:1998bc, Witten:1998qj}.
The holographic approach has been used to analyze the spectrum and structure of hadrons~\cite{deTeramond:2005su, Brodsky:2007hb, Branz:2010ub, Gutsche:2011vb, Li:2013oda, Gutsche:2017lyu, Lyubovitskij:2020gjz, Li:2021jqb, Zhang:2021itx, Chen:2022goa}, and also been successfully applied to the high energy scattering~\cite{Polchinski:2001tt, Polchinski:2002jw, Brower:2006ea, Hatta:2007he, Pire:2008zf, Domokos:2009hm, Domokos:2010ma, Marquet:2010sf, Watanabe:2012uc, Watanabe:2013spa, Watanabe:2015mia, Anderson:2016zon, Watanabe:2018owy, Xie:2019soz, Burikham:2019zbo, Watanabe:2019zny, Liu:2022out, Liu:2022zsa, Watanabe:2023rgp}.
The holographic approach in hadron physics is based on the conjecture that is dual between QCD and the string theory: the perturbative calculations of the string theory map the dynamics of the nonperturbative strong interaction in the higher dimensional curved spacetime.
The $S$-matrix in the Regge theory can be described by the bosonic strings, which is useful to investigate the scattering processes in the Regge regime, and it is known that the string amplitudes show the correct Regge behavior.
The glueball and mesons, which correspond to the Pomeron and Reggeon, are described in the closed and open string sector, respectively.
The intercept of the leading Pomeron trajectory is close to 1 but slightly greater.
On the other hand, the intercept of the Reggeon trajectory is less than 1.

In the preceding work~\cite{Liu:2022out}, the elastic $\pi p$ and $\pi\pi$ scattering were studied in holographic QCD, only considering the Pomeron exchange contribution.
We extend it in this study, and take into account both the Pomeron and Reggeon contribution, which is basically required to correctly describe the cross sections in the medium energy region.
In the present study the Pomeron and Reggeon exchange are described by the Reggeized $2^{++}$ glueball and vector meson propagator, respectively.
For the Pomeron-hadron couplings the gravitational form factors, which can be obtained with the bottom-up AdS/QCD models~\cite{Abidin:2008hn,Abidin:2009hr}, are employed.
We derive the expressions for the total and differential cross sections, and then numerically evaluate those.

Our model involves several parameters in total, which shall be determined with the experimental data.
However, by virtue of the universality of the Pomeron and Reggeon, the parameter values obtained in the preceding works on the proton-proton ($pp$) and proton-antiproton ($p \bar{p}$) scattering~\cite{Xie:2019soz, Liu:2022zsa} can directly be used in the present analysis.
The Pomeron-pion coupling constant was determined in Ref.~\cite{Liu:2022out}, but the used data have large uncertainties especially in the high energy region, which may have caused nonnegligible theoretical uncertainties.
Hence we newly determine it in this study.
Besides, the Reggeon-pion coupling constants are needed to be determined.
Since the charge difference affects the Reggeon couplings, we need to determine the Reggeon-$\pi^+$ and Reggeon-$\pi^-$ coupling constants separately.
All these three adjustable parameters are determined with the experimental data of the $\pi p$ total cross sections.
We show that the currently available data can be well described within the model.

Once the parameters are fixed, the $\pi \pi$ total cross sections and the $\pi p$ and $\pi \pi$ differential cross sections can be predicted without any additional parameters.
We present that our predictions for the $\pi^+ p$ and $\pi^- p$ differential cross section are consistent with the data.
Furthermore, for both the total and differential cross sections, the energy dependence of the Pomeron and Reggeon contribution is discussed, considering the contribution ratios.
Our results presented in this paper can be tested with the data to be taken in the future.

The structure of this paper is as follows.
In Sec.~\ref{sec:model} we explain the holographic description of the elastic $\pi p$ and $\pi \pi$ scattering in the Regge regime, taking into account the Pomeron and Reggeon exchange.
The expressions for the total and differential cross sections are derived.
We present our numerical results for the cross sections in Sec.~\ref{section3}, and give the conclusion of this work in Sec.~\ref{sec:conclusion}.

%%%%%%%%%%%%%%%%%%%%%%%%%%%%%%%%%%%%%%%%%%%%%%%%%%%%%%%%%%%%%%%%%%%%%%%%%%%%%%%
\section{Model setup}
\label{sec:model}
\begin{figure}
\centering
\begin{tabular}{ccccc}
\includegraphics[width=0.3\textwidth]{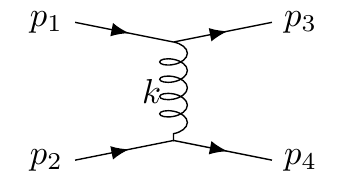}
\includegraphics[width=0.3\textwidth]{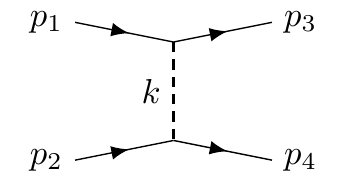}
\end{tabular}
\caption{
The left and right Feynman diagrams represent the $\pi p$($\pi \pi$) scattering with the $2^{++}$ glueball and vector meson exchange, respectively.
$p_1, p_2$ and $p_3, p_4$ are the initial and final four-momenta in the $t$-channel, respectively.
$k$ is the momentum transfer.
}
\label{Feynman}
\end{figure}

%%%%%%%%%%%%%%%%%%%%%%%%%%%%%%%%%%%%%%%%%%%%%%%%%%%%%%%%%%%%%%%%%%%%%%%%%%%%%%%
In this section the expressions for the total and differential cross sections of the elastic $\pi p$ and $\pi \pi$ scattering are derived, taking into account the Pomeron and Reggeon exchange.
The corresponding Feynman diagrams are shown in Fig.~\ref{Feynman}.
The Pomeron-hadron couplings are determined by the lowest state on the Pomeron trajectory, which is assumed as $2^{++}$ glueball.
The Reggeon-hadron couplings are determined by the lowest state on the Reggeon trajectory, which is assumed as the vector meson.
Since these two contributions can be considered separately, the total amplitudes for the $\pi p$ and $\pi \pi$ scattering can be written as
\begin{equation}
\mathcal{A}_{\rm tot}^{\pi p (\pi \pi)} = \mathcal{A}_{ g}^{\pi p (\pi \pi)} + \mathcal{A}_{ v}^{\pi p (\pi \pi)},
\end{equation}
where $\mathcal{A}_g$ and $\mathcal{A}_v$ represent amplitudes for the Pomeron and Reggeon exchange, respectively.
The upper index $\pi$ stands for $\pi^+$ or $\pi^-$.

The $2^{++}$ glueball field is described as a second-rank symmetric traceless tensor $h_{\mu\nu}$, assuming that the coupling of $h_{\mu\nu}$ to the QCD energy momentum tensor $T^{\mu\nu}$ is described by the action:
\begin{equation}
S = \lambda \int d^{4} x h_{\mu\nu} T^{\mu\nu},
\end{equation}
where $\lambda$ is a respective coupling.
The singlet state corresponds to the graviton, and the vertex of the glueball-proton-proton can be extracted from the matrix element of the energy momentum tensor $\langle p',s' | T^{\mu\nu} |~p,~s \rangle$, which can be expanded with three terms including the gravitational form factors~\cite{Pagels:1966zza} and can be written as
\begin{align}
\label{em}
\langle p',s'|T^{\mu\nu}|p,s\rangle =
\bar{u}(p',s')\bigg[&A_{p}(t)\frac{\gamma^{\mu}P^{\nu}_p + \gamma^{\nu}P^{\mu}_p}{2} \nonumber \\
&+B_{p} (t)\frac{ik_{p\rho} (P^{\mu}_p\sigma^{\nu\rho} + P^{\nu}_p\sigma^{\mu\rho})}{4m_{p}} \nonumber \\
&+C_{p}(t)\frac{k^{\mu}_pk^{\nu}_p-\eta^{\mu\nu}k_p^2}{m_{p}}\bigg] u(p,s),
\end{align}
where $\bar{u}(p',s')$ and $u(p,s)$ are the nucleon spinors, $m_p$ is the proton mass, the momentum transfer $k_p=p'-p=p_3-p_1$, $P_p=(p_1+p_3)/2=(p_2+p_4)/2$, $\eta^{\mu \nu}$ is the Minkowski matrix tensor, and $\sigma^{\mu \nu} = i [ \gamma^\mu, \gamma^\nu ] / 2$.
Since in the Regge regime $p_2\approx p_4$, we define $P_p=p_p$, where $p_p$ is the proton four-momentum of the initial or final state.
In the above equation, $A_p(t)$, $B_p(t)$ and $C_p(t)$ are the gravitational form factors of the proton.
The contribution of the $B_p(t)$ involved term is negligible when the momentum transfer is quite small.
Since the proton is a fermion and its wave function needs to satisfy the Dirac equation, the contribution of the $C_p(t)$ involved term vanishes.
Similarly, the vertex of the glueball-pion-pion can be extracted from the energy momentum tensor matrix element for the pion expanded in terms of two gravitational form factors $A_\pi (t)$ and $C_\pi (t)$:
\begin{equation}
\langle \pi^a(p_2)|T^{\mu\nu}|\pi^b(p_1)\rangle =
\delta^{ab}\bigg[2A_{\pi}(t)p^{\mu}_{\pi}p^{\nu}_{\pi} + \frac{1}{2}C_{\pi}(t)\big(k_{\pi}^2\eta^{\mu\nu} - k^{\mu}_{\pi}k^{\nu}_{\pi}\big)\bigg],
\end{equation}
where $p_{\pi}$ is the pion four-momentum.
The contribution of the $C_{\pi}(t)$ involved term is negligible in the Regge regime.
In our numerical evaluations, which will be presented in the next section, we employ the results obtained with the bottom-up AdS/QCD models~\cite{Abidin:2008hn,Abidin:2009hr} to specify $A_p(t)$ and $A_{\pi}(t)$.

Focusing on the Regge regime, the glueball-proton-proton vertex for the elastic scattering can be written as
\begin{equation}
\Gamma_{gpp}^{\mu\nu} =
\frac{i\lambda_{gpp}A_{p}(t)}{2}(\gamma^{\mu}p_{p}^{\nu} + \gamma^{\nu}p^{\mu}_{p}),
\end{equation}
where $\lambda_{gpp}$ is the coupling constant.
Likewise, the glueball-pion-pion vertex can be written as
\begin{equation}
\Gamma_{g\pi\pi}^{\mu\nu} = 
2i\lambda_{g\pi\pi}A_{\pi}(t)p_{\pi}^{\mu}p_{\pi}^{\nu},
\end{equation}
where $\lambda_{g\pi\pi}$ is the coupling constant.
The propagator of the massive $2^{++}$ glueball is expressed as~\cite{Yamada:1982dx}
\begin{equation}
D_{\alpha\beta\gamma\delta}^{g} =
- i \frac{d_{\alpha\beta\gamma\delta}(k)}{k^2+m_{g}^2},
\end{equation}
in which $m_g$ is the glueball mass, and $d_{\alpha\beta\gamma\delta}(k)$ is explicitly written as
\begin{align}
d_{\alpha\beta\gamma\delta} =
&\frac{1}{2}(\eta_{\alpha\gamma}\eta_{\beta\delta} + \eta_{\alpha\delta}\eta_{\beta\gamma}) - \frac{1}{2m_{g}^2}(k_{\alpha}k_{\delta}\eta_{\beta\gamma} + k_{\alpha} k_{\gamma}\eta_{\beta\delta} + k_{\beta}k_{\delta}\eta_{\alpha\gamma} + k_{\beta}k_{\gamma}\eta_{\alpha\delta}) \nonumber \\
&+\frac{1}{24}\left[\left(\frac{k^2}{m_{g}^2}\right)^2 - 3\left(\frac{k^2}{m_{g}^2}\right) - 6\right]\eta_{\alpha\beta}\eta_{\gamma\delta} - \frac{k^2 - 3m_{g}^2}{6m_{g}^4} (k_{\alpha}k_{\beta}\eta_{\gamma\delta} + k_{\gamma}k_{\delta}\eta_{\alpha\beta}) \nonumber \\
&+\frac{2k_{\alpha}k_{\beta}k_{\gamma}k_{\delta}}{3m_{g}^4}.
\end{align}
The amplitude of the elastic $\pi p$ scattering with the $2^{++}$ glueball exchange is given by
\begin{equation}
\mathcal{A}_{g}^{\pi p} =
\Gamma_{g\pi\pi}^{\alpha\beta}\bar{u}_{4}\Gamma_{gpp}^{\gamma \delta} u_{2}D_{\alpha\beta\gamma\delta}^{g}.
\end{equation}

The vertices of the vector-proton-proton and vector-pion-pion can be expressed as
\begin{align}
&\Gamma_{vpp}^{\mu}=-i\lambda_{vpp}\gamma^{\mu}, \\
&\Gamma_{v\pi\pi}^{\nu}=-2i\lambda_{v\pi\pi}p_{\pi}^{\nu},
\end{align}
respectively.
The propagator of the vector meson is given by~\cite{Anderson:2016zon}
\begin{equation}
D_{\mu\nu}^{v}(k) =
i \frac{\eta_{\mu\nu}}{k^2+m_{v}^2},
\end{equation}
where $m_v$ is the vector meson mass.
Hence the amplitude of the elastic $\pi p$ scattering with the vector meson exchange is written as
\begin{equation}
\mathcal{A}_{v}^{\pi p} =
\Gamma_{v\pi\pi}^{\nu}\bar{u}_3\Gamma_{vpp}u_1D_{\mu\nu}^{v}(k).
\end{equation}
With these equations, the total amplitude of the elastic $\pi p$ scattering is expressed as
\begin{align}
\mathcal{A}_{\mathrm{tot}}^{\pi p} =
&-i\lambda_{g\pi\pi}\lambda_{gpp}A_{\pi}(t)A_{p}(t)p_{\pi\mu}\bar{u}_4\gamma^{\mu}u_2\times\frac{1}{t-m_{g}^2} \nonumber \\
&+2i\lambda_{v\pi\pi}\lambda_{vpp}p_{\pi\nu}\bar{u}_3\gamma^{\nu}u_1\times\frac{1}{t-m_{v}^2}.
\end{align}
Similarly, the total amplitude of the elastic $\pi\pi$ scattering can be obtained as
\begin{align}
\mathcal{A}_{\mathrm{tot}}^{\pi\pi} =
- i \lambda_{gpp}^2A_\pi^2(t)s^2\times\frac{1}{t-m_g^2} + 2i\lambda_{v\pi\pi}^2s\times\frac{1}{t-m_v^2}.
\end{align}
Then, the total invariant amplitudes of the elastic $\pi p$ and $\pi \pi$ scattering are derived as
\begin{equation}\label{amplitude_pip}
\mathcal{A}_{\mathrm{tot}}^{\pi p}= - \lambda_{g\pi\pi}\lambda_{gpp}A_{\pi}(t)A_{p}(t)s^2\times\frac{1}{t-m_{g}^2} + 2\lambda_{v\pi\pi}\lambda_{vpp}s\times\frac{1}{t-m_{v}^2},
\end{equation}
\begin{equation}\label{amplitude_pipi}
\mathcal{A}_{\mathrm{tot}}^{\pi \pi}= - \lambda_{g\pi\pi}^2A_\pi^2(t)s^2\times\frac{1}{t-m_g^2} + 2\lambda_{v\pi\pi}^2s\times\frac{1}{t-m_v^2},
\end{equation}
respectively.

In the equations introduced above, only the lightest states on the Pomeron and Reggeon trajectory are considered, and taking into account the excited states of the strings is needed to include the higher spin states.
The excited states of the open and closed string correspond to the higher spin states which lie on the Reggeon and Pomeron trajectory, respectively.
Following the Reggeization procedure explained in detail in Ref.~\cite{Anderson:2016zon}, the bosonic open string four-tachyon amplitude can be written as
\begin{equation}
\mathcal{A}_o^4 (s,t,u) =
\widetilde{\mathcal{A}}_o(s,t) + \widetilde{\mathcal{A}}_o(u,t) + \widetilde{\mathcal{A}}_o(s,u).
\end{equation}
$\widetilde{\mathcal{A}}_o(x,y)$ is the Veneziano amplitude, which describes the string scattering in the flat space and is given by
\begin{equation}
\widetilde{\mathcal{A}}_o(x,y) =
iC\frac{\Gamma[-a_o(x)]\Gamma[-a_o(y)]}{\Gamma[-a_o(x)-a_o(y)]},
\end{equation}
where $a_o(x)$ is the trajectory of the open string.
For the Reggeon to be represented as the excited states, we replace $a_o(x)$ with $\alpha_R(x)$, which is the Reggeon trajectory and expressed as $\alpha_R(x)=\alpha_R(0)+\alpha_R'x$.
$\widetilde{\mathcal{A}}_o(s,u)$ has no contribution to the amplitude due to that it has no pole in the $t$-channel.
The poles of $\widetilde{\mathcal{A}}_o(s,t)$ and $\widetilde{\mathcal{A}}_o(u,t)$ in the $t$-channel are $a_o(t)=n$.
Expanding them around the poles, one finds
\begin{align}
&\widetilde{\mathcal{A}}_o(s,t) \approx i C e^{-i\pi a_o(t)}(\alpha'_os)^{a_o(t)}\Gamma[-a_o(t)], \\
&\widetilde{\mathcal{A}}_o(u,t) \approx i C (\alpha'_os)^{a_o(t)}\Gamma[-a_o(t)],
\end{align}
in which the open string amplitudes show the correct Regge behavior ($\mathcal{A} \approx s^{J}$, where $J$ is the spin of the exchange particles).
For the odd spin states, taking the difference between $\widetilde{\mathcal{A}}_o(s,t)$ and $\widetilde{\mathcal{A}}_o(u,t)$ represents the amplitude with the Reggeon exchange, which can be written as
\begin{equation}
\mathcal{A}_o^4 =
\widetilde{\mathcal{A}}_o(u,t) - \widetilde{\mathcal{A}}_o(s,t).
\end{equation}
This can be expanded around the pole $\alpha_R(t) = 1$ as
\begin{equation}
\mathcal{A}_R^4 =
i C \big(1-e^{-i\pi\alpha_v(t)}\big) (\alpha'_Rs)^{\alpha_R(t)}\Gamma[-\alpha_R(t)].
\end{equation}
Comparing this equation to the second term in the right-hand side of Eq.~\eqref{amplitude_pip}, one obtains
\begin{equation}
\frac{1}{t-m_v^2} \rightarrow
\alpha'_R e^{-\frac{i\pi\alpha_R(t)}{2}} \sin\bigg(\frac{\pi\alpha_R(t)}{2}\bigg) (\alpha'_Rs)^{\alpha_R(t) - 1}\Gamma[-\alpha_R(t)].
\end{equation}

The higher spin states on the Pomeron trajectory can be described as the excited states of the bosonic closed string, and the four-tachyon amplitude is given by
\begin{equation}
\mathcal{A}_c^4(s,t,u) =
2 \pi C \frac{\Gamma\big[ -\frac{a_c(t)}{2}\big] \Gamma\big[-\frac{a_c(s)}{2} \big] \Gamma \big[-\frac{a_c(u)}{2} \big]}{\Gamma \big[-\frac{a_c(s)}{2} - \frac{a_c(t)}{2}\big] \Gamma \big[-\frac{a_c(s)}{2}-\frac{a_c(u)}{2} \big] \Gamma \big[-\frac{a_c(t)}{2} - \frac{a_c(u)}{2}\big]},
\end{equation}
where $a_c(x) = 2 + \alpha'_{P} x/2$ is the Regge trajectory of the closed string, and $\alpha'_{P}/2$ is the slope of the closed string trajectory.
The square of the closed string mass is $m_c^2=-4/\alpha'_{P}$, and then $s+t+u = 4m_c^2 = -16/\alpha'_P$, which leads to $a_c(s)+a_c(t)+a_c(u) = -2$.
The amplitude is expanded around the pole $a_c(t)=2$ in the Regge regime as
\begin{equation}\label{v}
\mathcal{A}_c^4(s,t)\approx
2 \pi C e^{-\frac{i\pi a_c(t)}{2}} \bigg( \frac{\alpha'_Ps}{4}\bigg)^{a_c(t)}\frac{\Gamma\big[-\frac{a_c(t)}{2}\big]}{\Gamma\big[1 + \frac{a_c(t)}{2}\big]}.
\end{equation}
Since the $2^{++}$ glueball represents the lowest state of the Pomeron, $a_c(t)$ needs to be replaced with $\alpha_P(t)-2$.
Since the vertex is dimensionless in the string theory, comparing this equation to the right-hand side of Eq.~\eqref{amplitude_pip}, one obtains
\begin{equation}\label{g}
\frac{1}{t-m_g^2}\rightarrow\frac{\alpha'_P}{2}e^{-\frac{i\pi\alpha_P(t)}{2}}\bigg(\frac{\alpha'_Ps}{2}\bigg)^{\alpha_P(t)-2}\frac{\Gamma\big[3-\frac{\chi}{2}\big]\Gamma\big[1-\frac{\alpha_P(t)}{2}\big]}{\Gamma\big[2-\frac{\chi}{2}+\frac{\alpha_P(t)}{2}\big]},
\end{equation}
where $\chi = \alpha_P(s) + \alpha_P(u) + \alpha_P(t)$.

With the Reggeized propagators introduced above, the invariant amplitudes for the elastic $\pi p$ and $\pi\pi$ scattering are rewritten as
\begin{align}
\mathcal{A}_{\mathrm{tot}}^{\pi p} =
&- \lambda_{g\pi\pi}\lambda_{gpp} A_{\pi}(t) A_{p}(t) s^2\frac{\alpha'_P}{2}e^{-\frac{i\pi\alpha_P(t)}{2}}\bigg( \frac{\alpha'_Ps}{2}\bigg)^{\alpha_P(t) - 2}\frac{\Gamma\big[3 - \frac{\chi}{2}\big]\Gamma\big[1-\frac{\alpha_P(t)}{2}\big]}{\Gamma\big[2 - \frac{\chi}{2} + \frac{\alpha_P(t)}{2}\big]} \nonumber \\
&+ 2 \lambda_{v\pi\pi}\lambda_{vpp} s \alpha'_R e^{-\frac{i\pi\alpha_R(t)}{2}} \sin \bigg( \frac{\pi\alpha_R(t)}{2}\bigg)(\alpha'_R s)^{\alpha_R(t) - 1}\Gamma[-\alpha_R(t)],
\end{align}
\begin{align}
\mathcal{A}_{\mathrm{tot}}^{\pi \pi} =
&- \frac{\lambda_{g\pi\pi}^2}{2}A_{\pi}^2 (t) s^2\alpha'_P e^{-\frac{i\pi\alpha_P(t)}{2}}\bigg( \frac{\alpha'_P s}{2}\bigg)^{\alpha_P(t) - 2}\frac{\Gamma\big[3 - \frac{\chi}{2}\big]\Gamma\big[1-\frac{\alpha_P(t)}{2}\big]}{\Gamma\big[2 - \frac{\chi}{2} + \frac{\alpha_P(t)}{2}\big]} \nonumber \\
&+ 2 \lambda_{v\pi\pi}^2 s \alpha'_R e^{-\frac{i\pi\alpha_R(t)}{2}}\sin\bigg( \frac{\pi\alpha_R(t)}{2}\bigg)(\alpha'_R s)^{\alpha_R(t) - 1}\Gamma[-\alpha_R(t)],
\end{align}
respectively.

Then, the differential cross section of the $\pi p$ scattering is derived as
\begin{align}
\frac{d\sigma^{\pi p}}{dt}&=
\frac{1}{16\pi s^2}|\mathcal{A}_{\mathrm{tot}}^{\pi p}|^2 \nonumber \\
&= \frac{1}{16\pi}\big(\lambda_{g\pi\pi}\lambda_{gpp}A_\pi(t)A_p(t) s\big)^2~\bigg[\frac{\alpha'_P}{2}\bigg(\frac{\alpha'_Ps}{2}\bigg)^{\alpha_P(t) - 2}\frac{\Gamma\big[3-\frac{\chi}{2}\big]\Gamma\big[1-\frac{\alpha_P(t)}{2}\big]}{\Gamma\big[2-\frac{\chi}{2} + \frac{\alpha_P(t)}{2}\big]}\bigg]^2 \nonumber \\
&~~~~~+ \frac{1}{4\pi}(\lambda_{v\pi\pi}\lambda_{vpp})^2~\bigg[\alpha'_R \sin \bigg(\frac{\pi\alpha_R(t)}{2}\bigg)(\alpha'_R s)^{\alpha_R(t) - 1}\Gamma[-\alpha_R(t)]\bigg]^2 \nonumber \\
&~~~~~- \frac{1}{8\pi}\big(\lambda_{g\pi\pi}\lambda_{gpp}\lambda_{v\pi\pi}\lambda_{vpp} A_\pi(t) A_p(t) s\big)\bigg[\frac{\alpha'_P}{2}e^{-\frac{i\pi\alpha_P(t)}{2}}\bigg(\frac{\alpha'_Ps}{2}\bigg)^{\alpha_P(t)-2}\frac{\Gamma\big[3 - \frac{\chi}{2}\big]\Gamma\big[1 - \frac{\alpha_P(t)}{2}\big]}{\Gamma\big[2 - \frac{\chi}{2} + \frac{\alpha_P(t)}{2}\big]}~, \nonumber \\
&~~~~~~~~~~~~~~~~~~~~~~~~~~~~~~~~~ \alpha'_v e^{-\frac{i\pi\alpha_R(t)}{2}}\sin\bigg(\frac{\pi\alpha_R(t)}{2}\bigg)(\alpha'_R s)^{\alpha_R(t) - 1}\Gamma[-\alpha_R(t)]\bigg]^*,
\end{align}
where $[x,y]^*=xy^*+x^*y$, and the asterisk indicates complex conjugation.
The three terms in the right-hand side represent the contributions of the Pomeron exchange, the Reggeon exchange and the cross term, respectively.
Similarly, the differential cross section of the $\pi \pi$ scattering can be obtained as
\begin{align}
\frac{d\sigma^{\pi\pi}}{dt} =
&\frac{1}{16\pi}\lambda_{g\pi\pi}^4 A_\pi^4(t) s^2~\bigg[\frac{\alpha'_P}{2}\bigg(\frac{\alpha'_P s}{2}\bigg)^{\alpha_P(t) - 2}\frac{\Gamma\big[3 - \frac{\chi}{2}\big]\Gamma\big[1 - \frac{\alpha_P(t)}{2}\big]}{\Gamma\big[2 - \frac{\chi}{2} + \frac{\alpha_P(t)}{2}\big]}\bigg]^2 \nonumber \\
&+ \frac{1}{4\pi}\lambda_{v\pi\pi}^4~\bigg[\alpha'_R \sin \bigg(\frac{\pi\alpha_R(t)}{2}\bigg)(\alpha'_R s)^{\alpha_R(t) - 1}\Gamma[-\alpha_R(t)]\bigg]^2 \nonumber \\
&- \frac{1}{8\pi}\big(\lambda_{g\pi\pi}^2\lambda_{v\pi\pi}^2 A_\pi^2 (t) s \big)\bigg[\frac{\alpha'_P}{2}e^{-\frac{i\pi\alpha_P(t)}{2}}\bigg(\frac{\alpha'_Ps}{2}\bigg)^{\alpha_P(t) - 2}\frac{\Gamma\big[3 - \frac{\chi}{2}\big]\Gamma\big[1 - \frac{\alpha_g (t)}{2}\big]}{\Gamma\big[2 - \frac{\chi}{2} + \frac{\alpha_P(t)}{2}\big]}~, \nonumber \\
&~~~~~~~~~~~~~~~~~~~~~~~~~~~~~~~~~~ \alpha'_v e^{-\frac{i\pi\alpha_R(t)}{2}}\sin\bigg(\frac{\pi\alpha_R(t)}{2}\bigg)(\alpha'_R s)^{\alpha_R (t) - 1}\Gamma[- \alpha_R (t)]\bigg]^*.
\end{align}

Applying the optical theorem, the total cross sections of the $\pi p$ and $\pi \pi$ scattering can be expressed as
\begin{align}
\sigma^{\pi p}_{\mathrm{tot}}&=
\frac{1}{s}\mathrm{Im}{\mathcal{A}^{\pi p}_{\mathrm{tot}}(s,t = 0)} \nonumber \\
&= \lambda_{g\pi\pi}\lambda_{gpp}\frac{\Gamma\big[3 - \frac{\chi}{2}\big]\Gamma\big[1 - \frac{\alpha_P(t)}{2}\big]}{\Gamma\big[2 - \frac{\chi}{2} + \frac{\alpha_P(t)}{2}\big]}\bigg(\frac{\alpha'_P s}{2}\bigg)^{\alpha_P (0) - 1} \sin \bigg(\frac{\pi\alpha_P (0)}{2}\bigg) \nonumber \\
&~~~~~-2 \lambda_{v\pi\pi}\lambda_{vpp}\alpha'_R \sin^2 \bigg(\frac{\pi\alpha_R (0)}{2}\bigg)(\alpha'_R s)^{\alpha_R (0) - 1}\Gamma[-\alpha_R (0)],
\end{align}
\begin{align}
\sigma^{\pi\pi}_{\mathrm{tot}} =
&\frac{1}{2}\lambda_{g\pi\pi}^2 \alpha'_P \frac{\Gamma\big[3 - \frac{\chi}{2}\big]\Gamma\big[1 - \frac{\alpha_P(t)}{2}\big]}{\Gamma\big[2 - \frac{\chi}{2} + \frac{\alpha_P(t)}{2}\big]}\bigg(\frac{\alpha'_P s}{2}\bigg)^{\alpha_P (0) - 2} \sin \bigg(\frac{\pi\alpha_P (0)}{2}\bigg) \nonumber \\
&- 2 \lambda_{v\pi\pi}^2 \alpha'_R \sin^2 \bigg(\frac{\pi\alpha_R (0)}{2}\bigg)(\alpha'_R s)^{\alpha_R (0) - 1}\Gamma[- \alpha_R (0)],
\end{align}
respectively.

%%%%%%%%%%%%%%%%%%%%%%%%%%%%%%%%%%%%%%%%%%%%%%%%%%%%%%%%%%%%%%%%%%%%%%%%%%%%%%%
\section{Numerical results}
\label{section3}
%%%%%%%%%%%%%%%%%%%%%%%%%%%%%%%%%%%%%%%%%%%%%%%%%%%%%%%%%%%%%%%%%%%%%%%%%%%%%%%
In this section we numerically evaluate the total and differential cross sections of the $\pi p$ and $\pi \pi$ scattering, whose analytical expressions are introduced in the previous section, and display the results.
The present model involves nine parameters in total.
However, by virtue of the universality of the Pomeron and Reggeon, for six of them we can employ the values determined in the preceding works.
For the three parameters $\left\{\alpha_P(0),\alpha'_P,\lambda_{gpp}\right\}$ we utilize the results obtained in Ref.~\cite{Xie:2019soz}, in which the $pp$ and $p \bar{p}$ cross sections were studied by only considering the Pomeron exchange, focusing on the high energy region.
For another three parameters $\left\{\alpha_R(0),\alpha'_R,\lambda_{vpp}\right\}$ we utilize the results obtained in Ref.~\cite{Liu:2022zsa}, in which both the Pomeron and Reggeon exchange were taken into account to investigate the $pp$ and $p \bar{p}$ cross sections in the medium energy region.
The parameter values taken from these preceding works are summarized in Table~\ref{table}.
\begin{table}[tb]
\centering
\caption{Parameter values.}
\begin{tabular}{l l l}
\hline
Parameter&~~~Value&~~~Source\\                
\hline
$\alpha_P(0)$ &~~~1.086 &~~~fit to $pp(p \bar{p})$ data at high energies~\cite{Xie:2019soz} \\
$\alpha'_P$ &~~~0.377~$\rm{GeV}^{-2}$ &~~~fit to $pp(p \bar{p})$ data at high energies~\cite{Xie:2019soz} \\
$\lambda_{gpp}$ &~~~9.699~$\rm{GeV}^{-1}$ &~~~fit to $pp(p \bar{p})$ data at high energies~\cite{Xie:2019soz} \\
$\alpha_R(0)$ &~~~0.444 &~~~fit to $pp(p \bar{p})$ data at medium energies~\cite{Liu:2022zsa} \\
$\alpha'_R$ &~~~0.925$~\rm{GeV}^{-2}$ &~~~fit to $pp(p \bar{p})$ data at medium energies~\cite{Liu:2022zsa} \\
$\lambda_{vpp}$ &~~~7.742~ &~~~fit to $pp(p \bar{p})$ data at medium energies~\cite{Liu:2022zsa} \\
$\lambda_{g\pi\pi}$ &~~~3.361~$\pm~0.002~\rm{GeV}^{-1}$ &~~~this work \\
$\lambda_{v\pi^{+}\pi^{+}}$ &~~~4.528~$\pm~0.023$ &~~~this work \\
$\lambda_{v\pi^{-}\pi^{-}}$ &~~~6.049~$\pm~0.022$ &~~~this work \\
\hline
\end{tabular}
\label{table}
\end{table}
The Pomeron-pion coupling constant $\lambda_{g \pi \pi}$ was determined in the previous work~\cite{Liu:2022out}, but the used data have large uncertainties especially in the high energy region, which may have caused nonnegligible theoretical uncertainties.
Hence, in this study we newly determine it with the experimental data.
Besides $\lambda_{g \pi \pi}$, the Reggeon-pion coupling constants are needed to be determined.
Since the charge difference affects the Reggeon couplings, we need to determine the Reggeon-$\pi^+$ and Reggeon-$\pi^-$ coupling constants, $\lambda_{v\pi^{+}\pi^{+}}$ and $\lambda_{v\pi^{-}\pi^{-}}$, separately.

\begin{figure}[tb]
\centering
\includegraphics[width=0.65\textwidth]{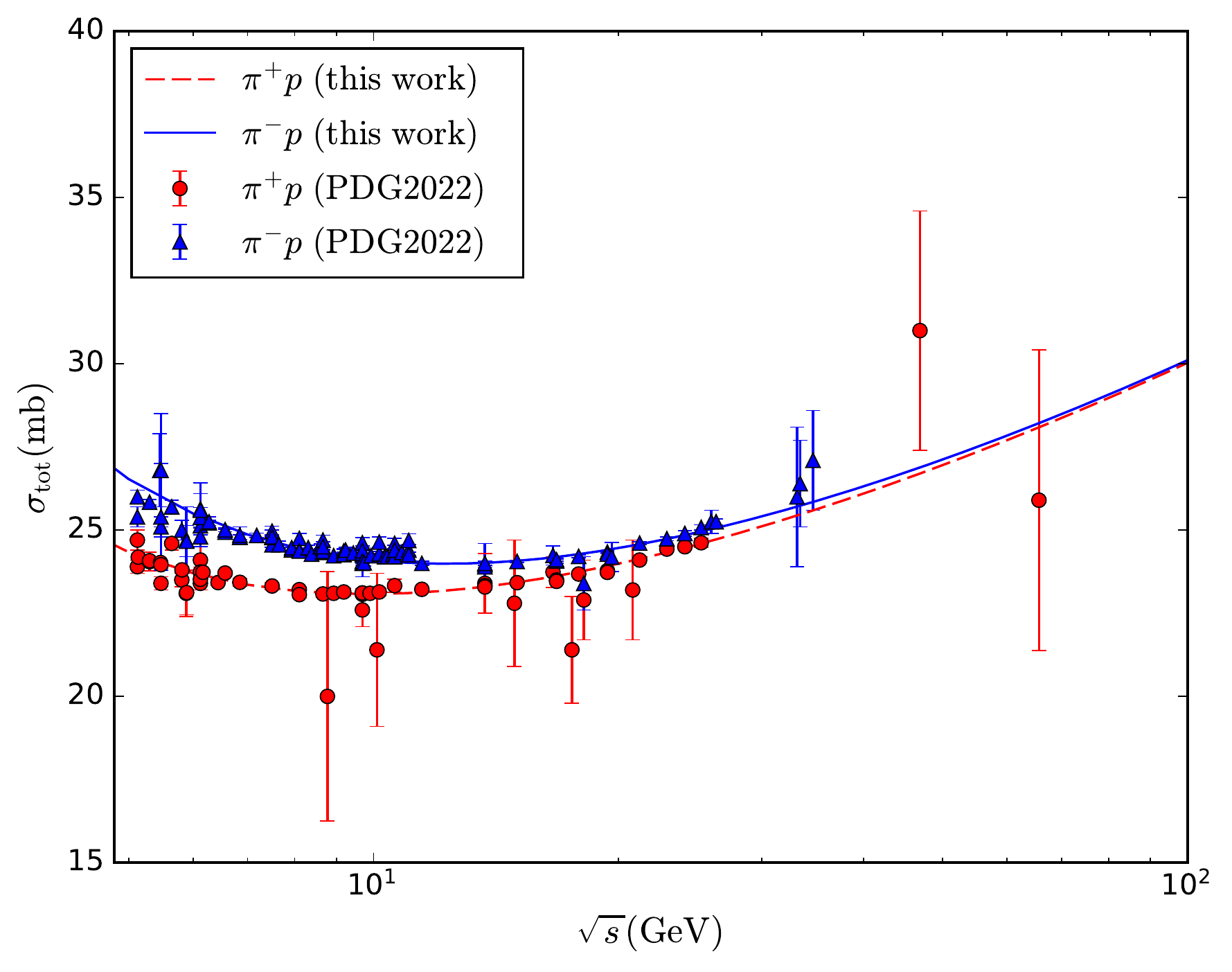}
\caption{
The $\pi p$ total cross sections as a function of $\sqrt{s}$.
The solid and dashed curves represent our calculations for the $\pi^- p$ and $\pi^+ p$ scattering, respectively.
The experimental data are taken from Ref.~\cite{ParticleDataGroup:2022pth}.
}
\label{TCS_low}
\end{figure}

\begin{figure}[tb]
\centering
\includegraphics[width=0.65\textwidth]{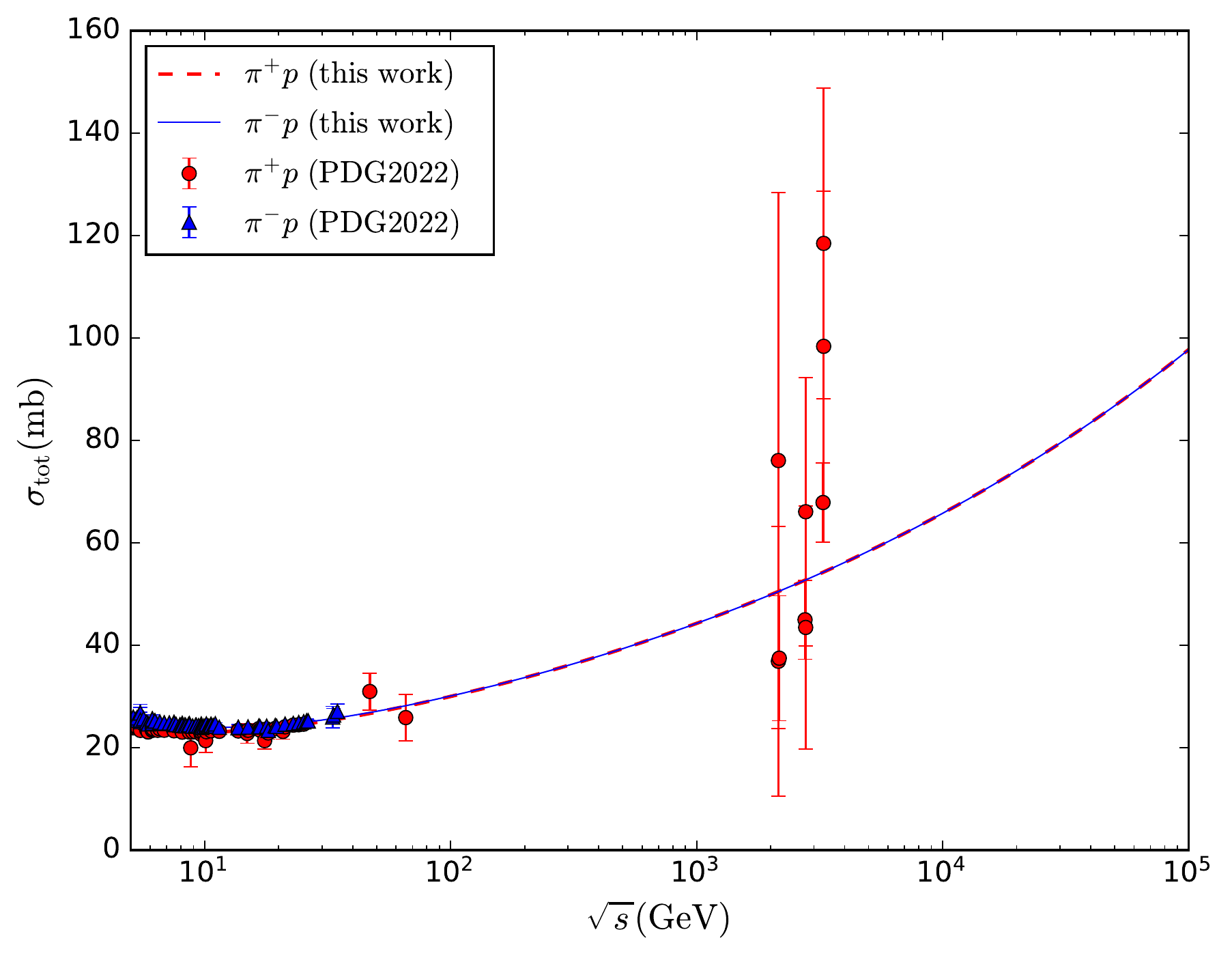}
\caption{
Similar to Fig.~\ref{TCS_low}, but for the wider $\sqrt{s}$ range.
}
\label{TCS_high}
\end{figure}

\begin{figure}[tb]
\centering
\includegraphics[width=0.65\textwidth]{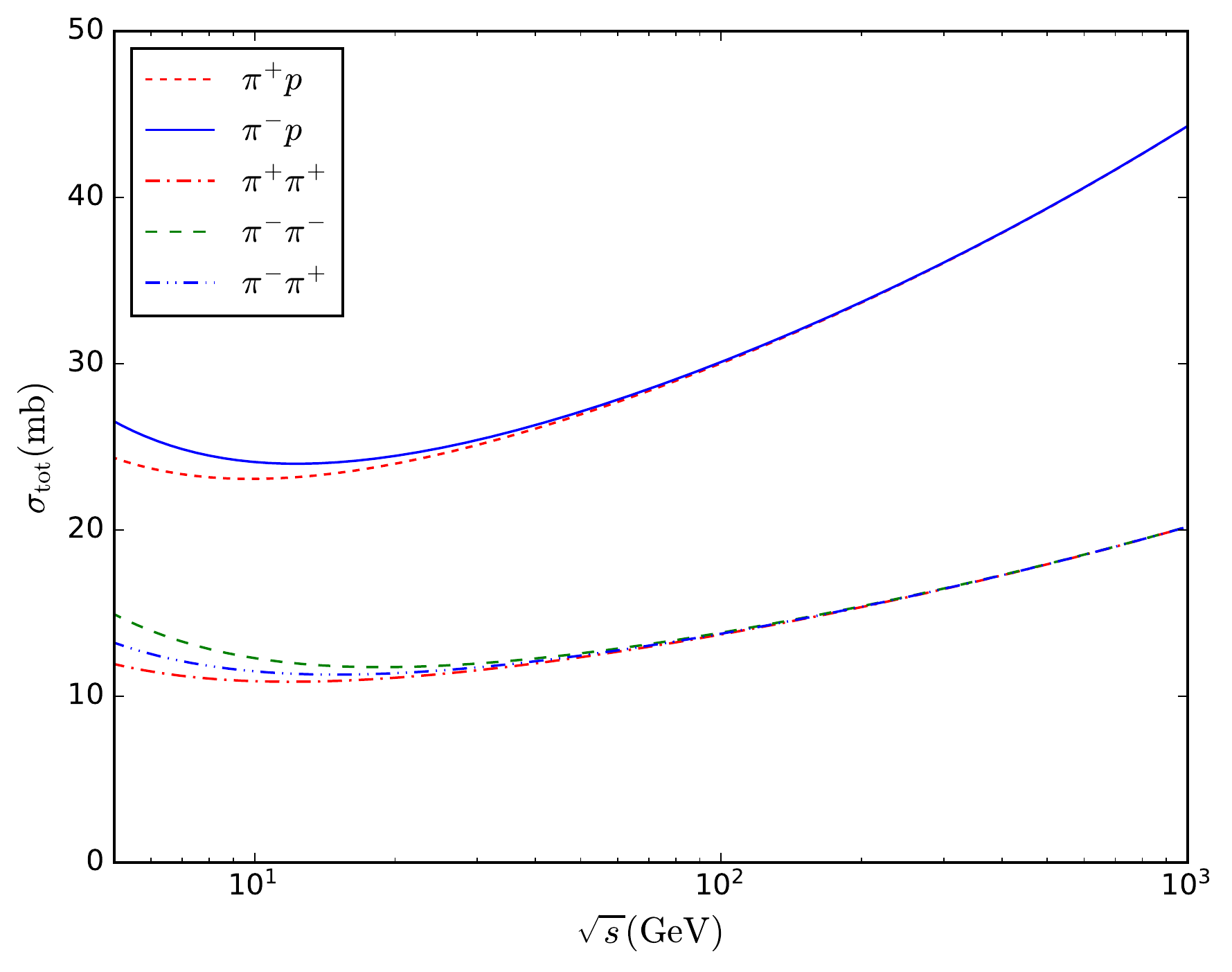}
\caption{
The $\pi p$ and $\pi \pi$ total cross sections as a function of $\sqrt{s}$.
}
\label{TCS_prediction}
\end{figure}

\begin{figure}[tb]
\centering
\begin{tabular}{cc}
\includegraphics[width=0.48\textwidth]{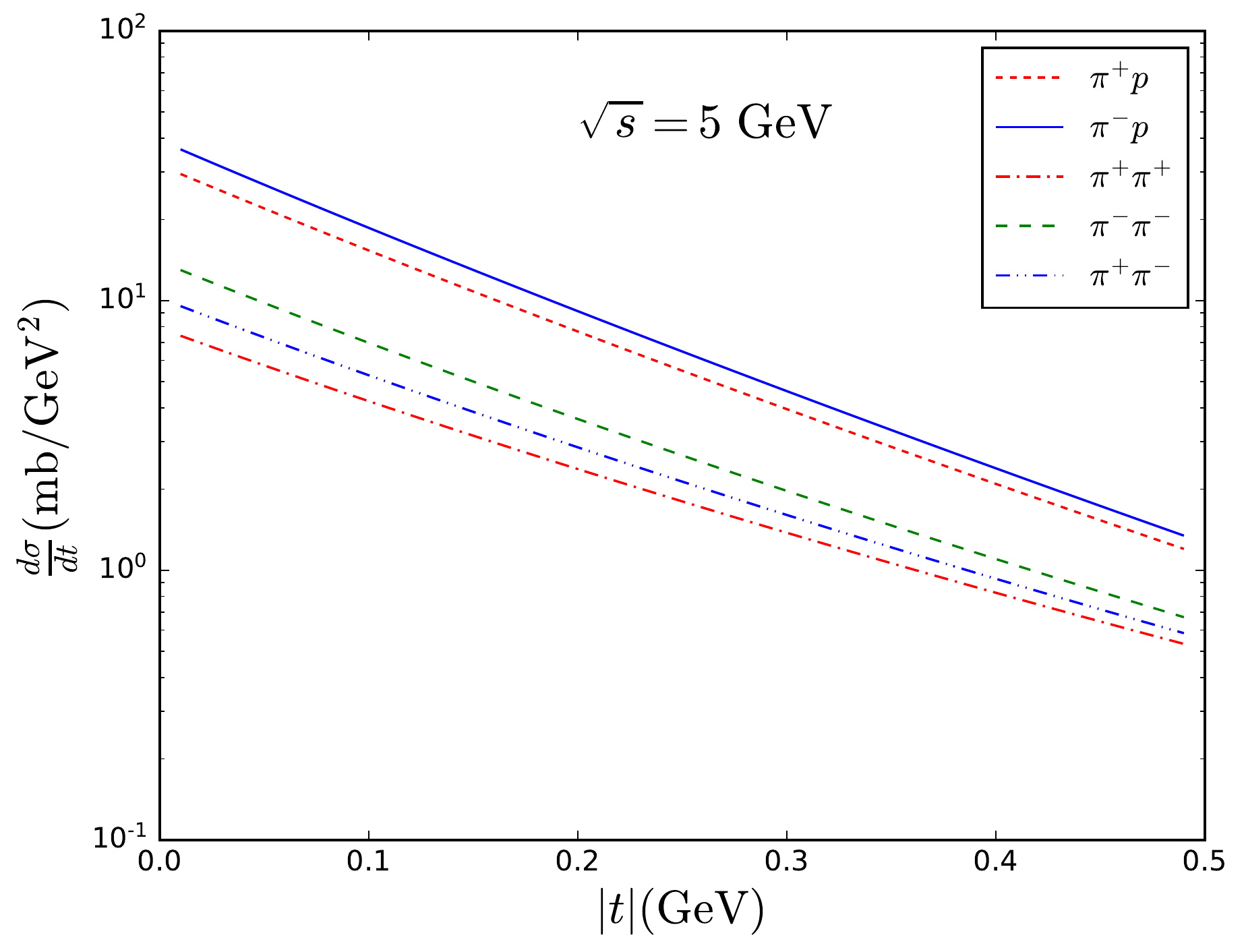}
\includegraphics[width=0.48\textwidth]{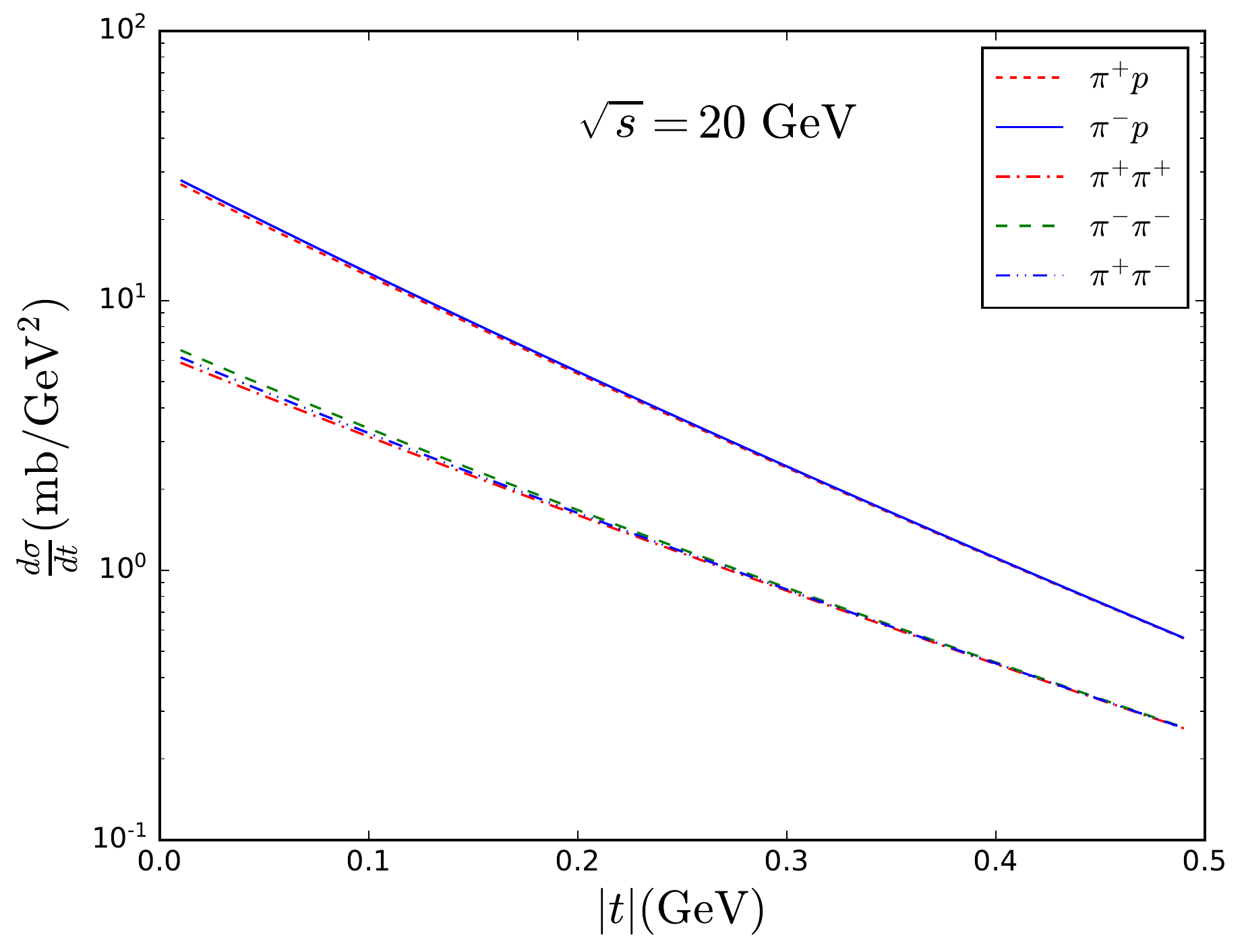}
\end{tabular}
\caption{
The $\pi p$ and $\pi \pi$ differential cross sections as a function of $|t|$ for $\sqrt{s} = 5$ and $20$~GeV.
}
\label{DCS_prediction}
\end{figure}

We determine the three adjustable parameters $\left\{\lambda_{g\pi\pi},\lambda_{v\pi^-\pi^-},\lambda_{v\pi^+\pi^+} \right\}$ by numerical fitting with the experimental data of the $\pi p$ total cross sections, focusing on the kinematic range, $5 \leq \sqrt{s} \leq 100$~GeV.
To perform this, the MINUIT package~\cite{James:1975dr} and the data summarized by the Particle Data Group (PDG) in 2022~\cite{ParticleDataGroup:2022pth} are utilized.
The obtained parameter values are shown in Table~\ref{table}, and the resulting $\pi p$ total cross sections are displayed in Fig.~\ref{TCS_low}.
It is seen from the figure that the both $\pi^+ p$ and $\pi^- p$ data are well described with the model in the whole considered kinematic region.
The newly determined value for $\lambda_{g \pi \pi}$ is smaller than that obtained in Ref.~\cite{Liu:2022out}.
Since this also affects the magnitudes of the total cross sections in the high energy region, the previous results shall be updated.
We show in Fig.~\ref{TCS_high}
our calculations in the wider $\sqrt{s}$ range.
Since the available data in the high energy region, in which $\sqrt{s} > 100$~GeV, are quite limited and the uncertainties of those are huge, more precise data are necessary especially in the TeV region to test our results.
Focusing on the high energy region, in which the Pomeron exchange contribution is dominant, the resulting total cross section ratios are found to be
\begin{equation}
\frac{\sigma_{\mathrm{tot}}^{\pi p}}{\sigma_{\mathrm{tot}}^{p p}} = 0.62,
\ \ \ \ \
\frac{\sigma_{\mathrm{tot}}^{\pi \pi}}{\sigma_{\mathrm{tot}}^{p p}} = 0.28.
\end{equation}
This ratio $\sigma_{\mathrm{tot}}^{\pi p} / \sigma_{\mathrm{tot}}^{p p}$ is consistent with the result obtained in Ref.~\cite{Watanabe:2018owy} and the value which can be extracted from the results presented in Ref.~\cite{Donnachie:1992ny}, although the other ratio $\sigma_{\mathrm{tot}}^{\pi \pi} / \sigma_{\mathrm{tot}}^{p p}$ is obviously smaller compared to the result in Ref.~\cite{Watanabe:2018owy}.
Once the three adjustable parameters are determined, the $\pi \pi$ total cross sections and the $\pi p$ and $\pi \pi$ differential cross sections can be calculated without any additional parameters.
We display our predictions for the total cross sections in Fig.~\ref{TCS_prediction}
and for the differential cross sections in Fig.~\ref{DCS_prediction}.
From these figures it is seen that the differences between $\pi^+$ and $\pi^-$ become smaller as $\sqrt{s}$ increases.

\begin{figure}[tb]
\centering
\includegraphics[width=1.0\textwidth]{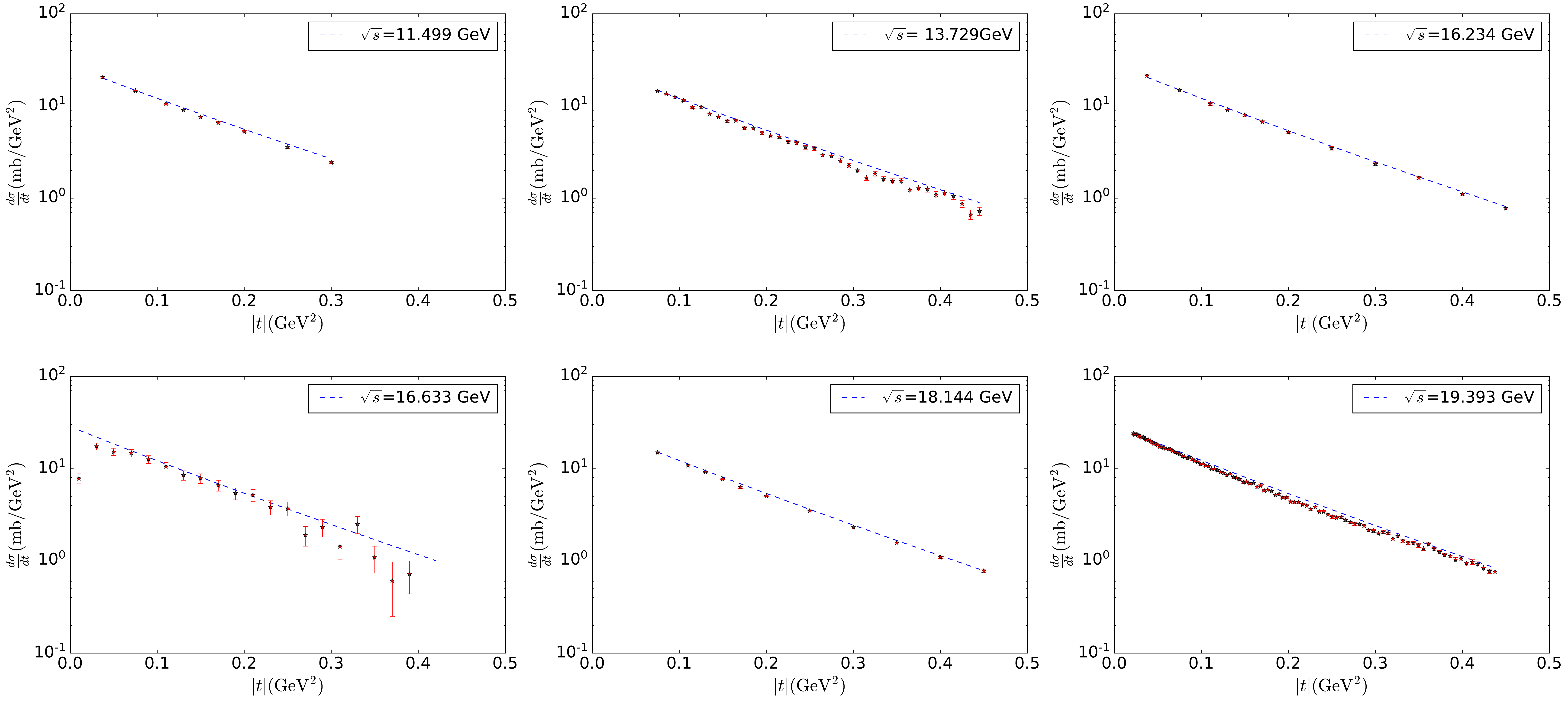}
\caption{
The $\pi^+ p$ differential cross section as a function of $|t|$ for $\sqrt{s} \geq 10$~GeV.
The dashed curves represent our predictions.
The experimental data are taken from Refs.~\cite{FermilabSingleArmSpectrometerGroup:1976krf,Schiz:1979rh,Akerlof:1976gk,Brick:1982dy}.
}
\label{DCS_pi+p}
\end{figure}

\begin{figure}[tb]
\centering
\begin{tabular}{cc}
\includegraphics[width=1\textwidth]{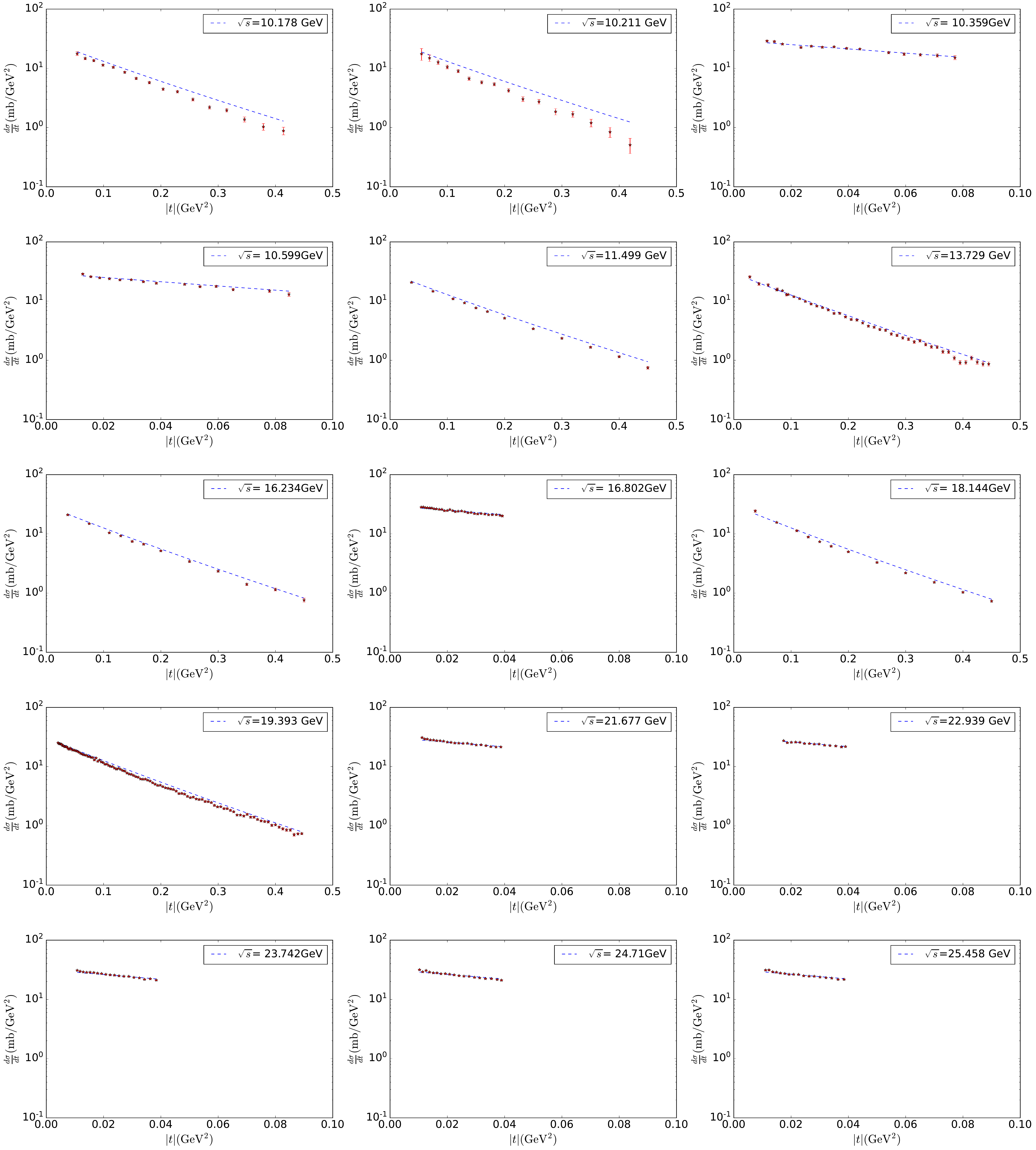}
\end{tabular}
\caption{
Similar to Fig.~\ref{DCS_pi+p}, but for the $\pi^- p$ scattering.
The experimental data are taken from Refs.~\cite{FermilabSingleArmSpectrometerGroup:1976krf, Burq:1982ja, Akerlof:1976gk, Derevshchikov:1973ik, Apokin:1975ap}.
}
\label{DCS_pi-p}
\end{figure}

Then, we present the comparisons between our predictions and the experimental data for the $\pi p$ differential cross sections.
To focus on the Regge regime and also to avoid the effect of the Coulomb scattering~\cite{Amos:1985wx, UA4:1987gcp}, we choose the data in the range, $\sqrt{s} \geq 10$~GeV and $0.01 \leq |t| \leq 0.45$~GeV$^2$, for the $\pi^+ p$~\cite{FermilabSingleArmSpectrometerGroup:1976krf,Schiz:1979rh,Akerlof:1976gk,Brick:1982dy} and $\pi^- p$~\cite{FermilabSingleArmSpectrometerGroup:1976krf, Burq:1982ja, Akerlof:1976gk, Derevshchikov:1973ik, Apokin:1975ap} scattering.
We show the results for the $\pi^+ p$ scattering in Fig.~\ref{DCS_pi+p}.
Although the $\sqrt{s}$ range, in which the data exist, is narrow, it is found that our predictions agree with the data in the whole considered kinematic region.
The results for the $\pi^- p$ scattering are displayed in Fig.~\ref{DCS_pi-p}.
In the first two panels for $\sqrt{s} = 10.178$ and $10.211$~GeV, we can find obvious deviations between our predictions and the data.
For these two the $|t|/s$ values are relatively larger compared to the others, which may be the possible reason.
It is seen from the other panels that our predictions are consistent with the data.

\begin{figure}
\centering
\begin{tabular}{cc}
\includegraphics[width=0.48\textwidth]{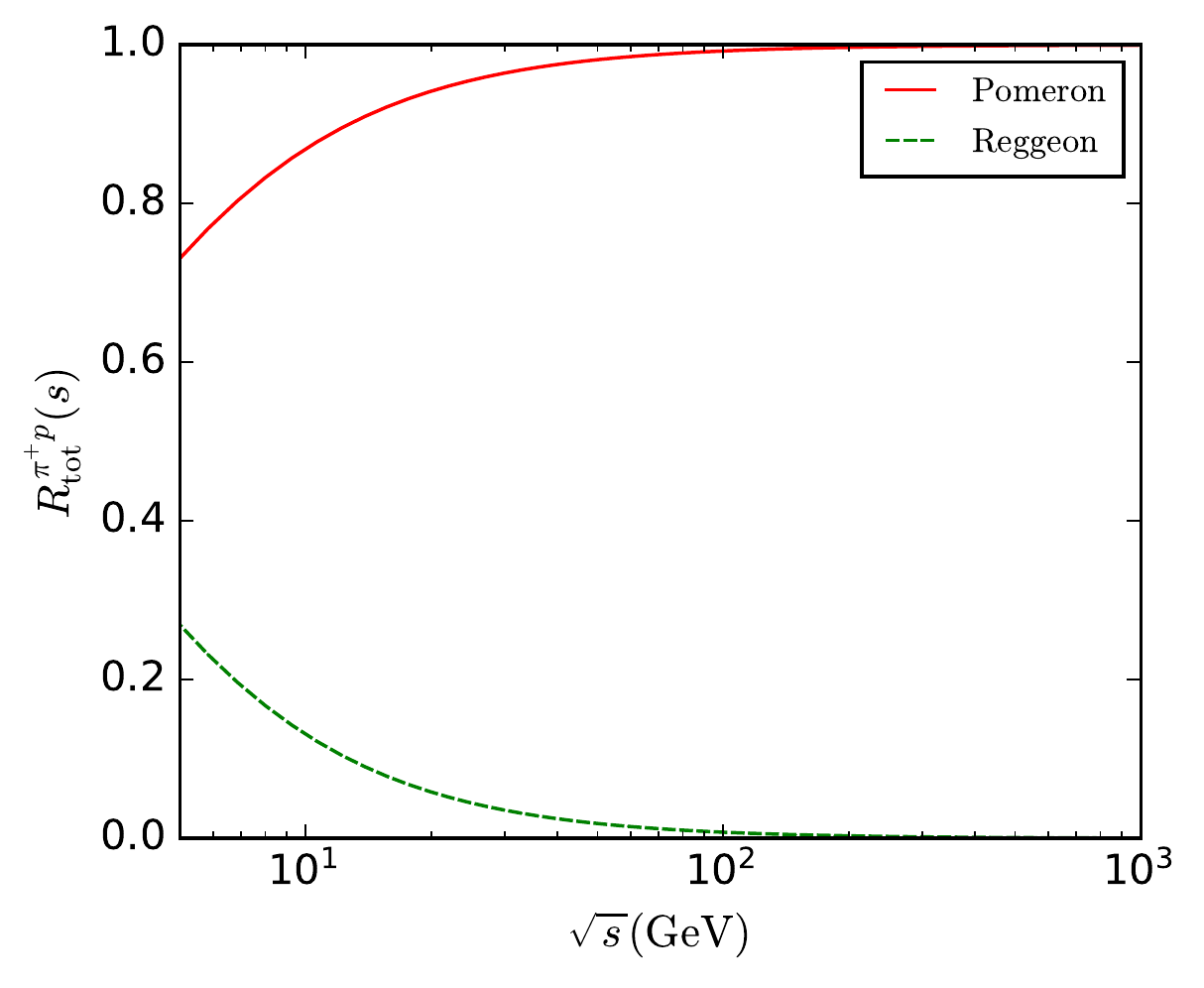}
\includegraphics[width=0.48\textwidth]{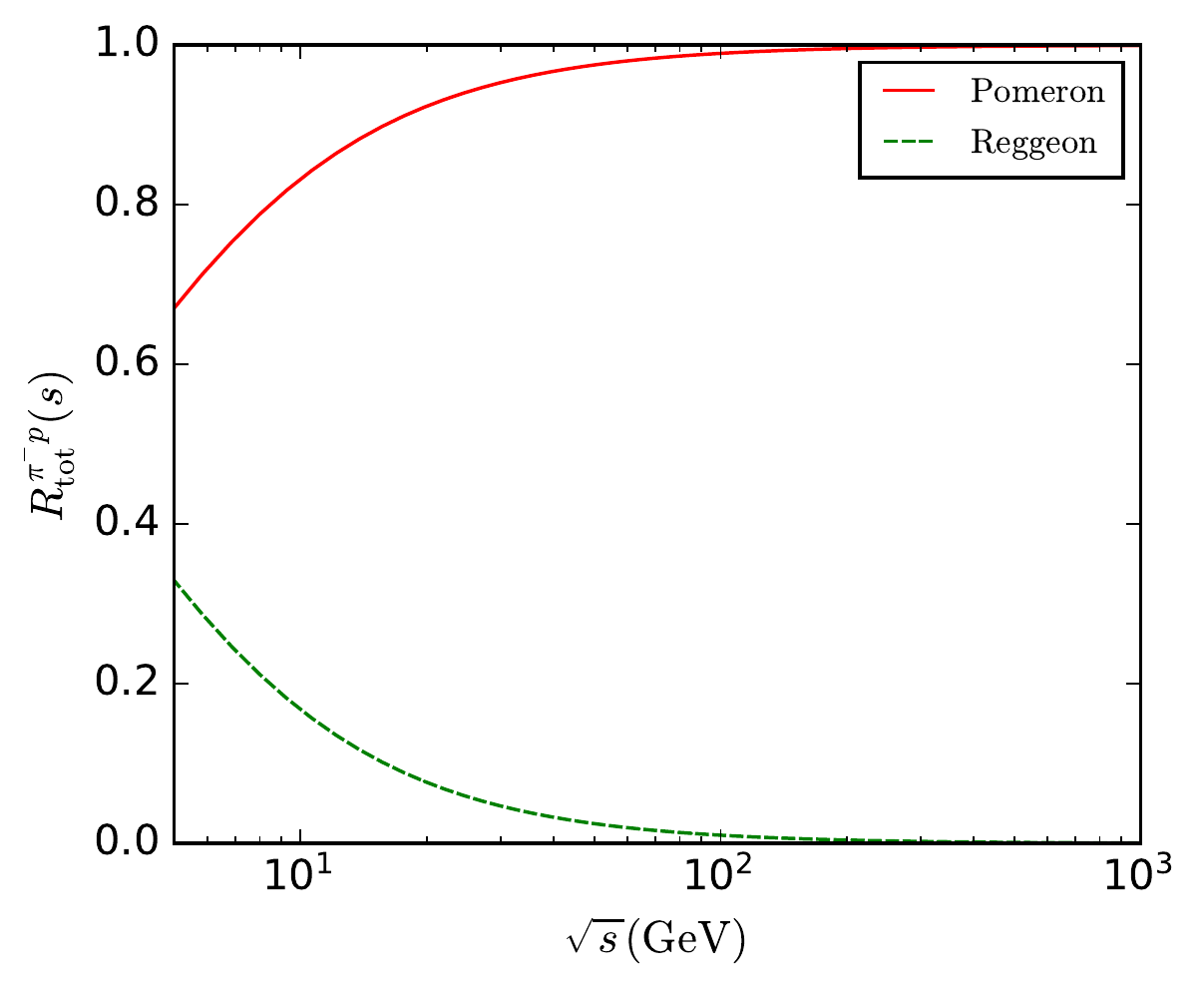}
\end{tabular}
\caption{
The contribution ratios for the $\pi^+p$ (left) and $\pi^-p$ (right) total cross section as a function of $\sqrt{s}$.
The solid and dashed curves represent the ratios for the Pomeron and Reggeon exchange, respectively.
}
\label{contribution_pip_tcs}
\end{figure}

\begin{figure}
\centering
\begin{tabular}{cc}
\includegraphics[width=0.48\textwidth]{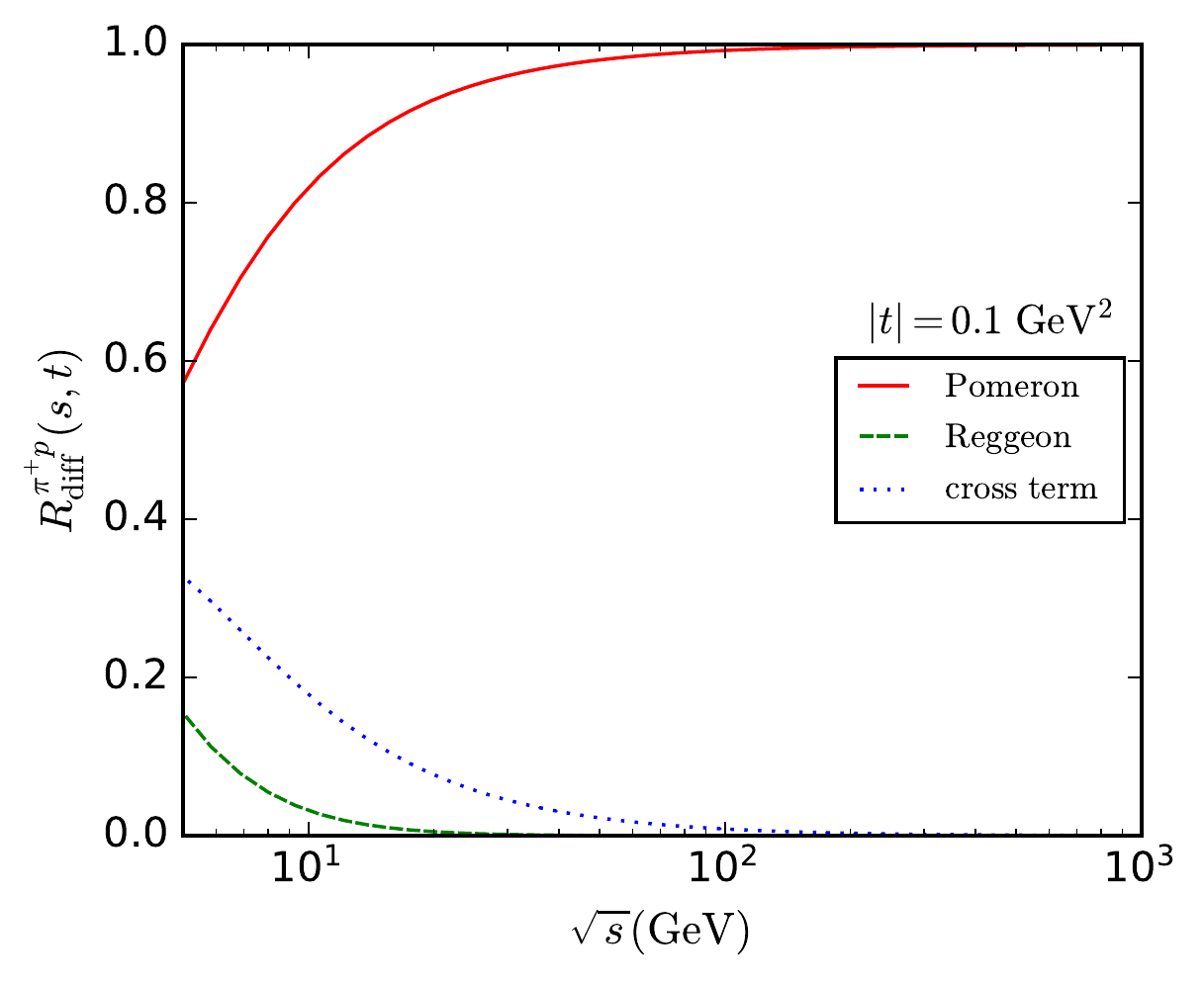}
\includegraphics[width=0.48\textwidth]{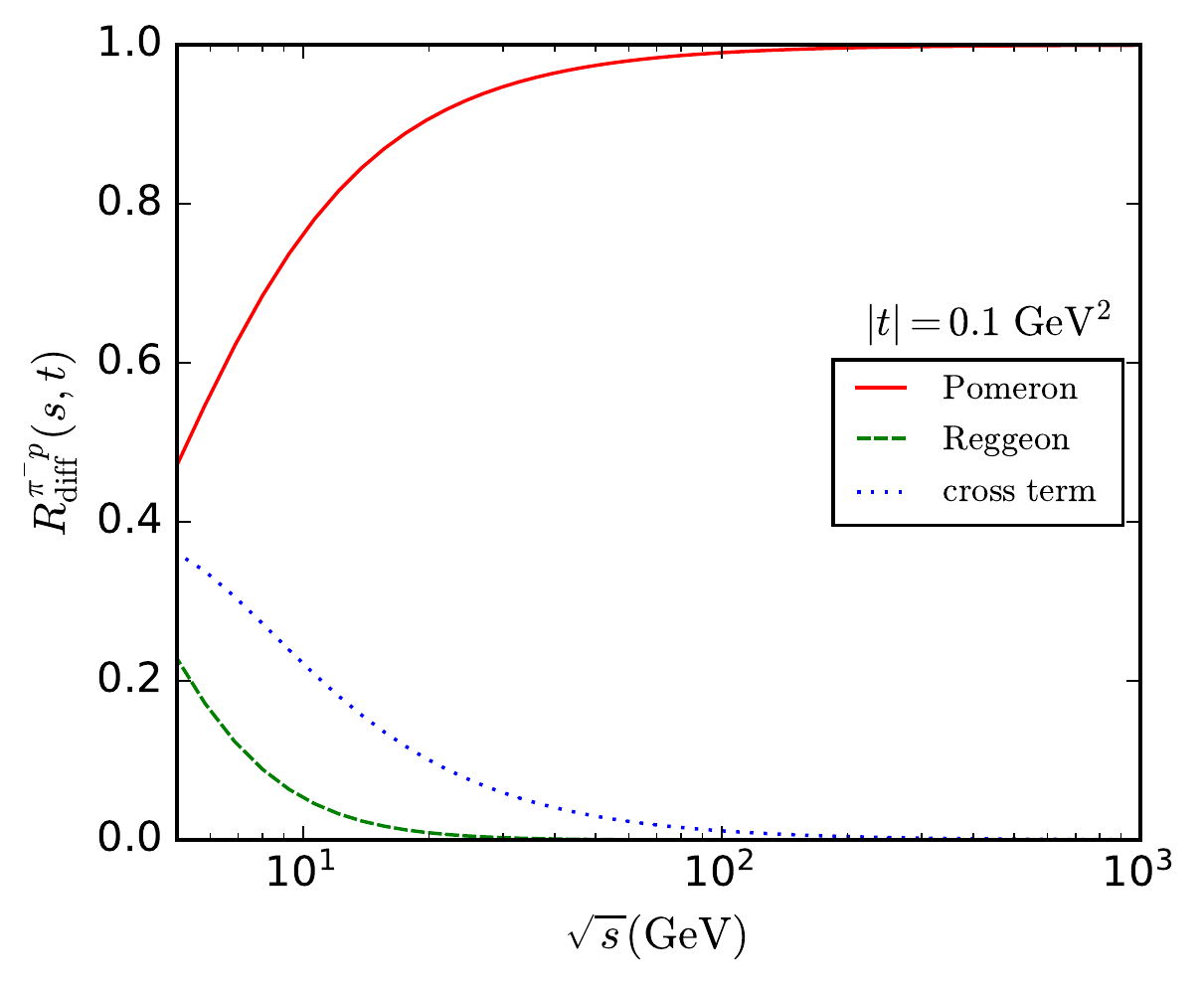}
\end{tabular}
\caption{
The contribution ratios for the $\pi^+p$ (left) and $\pi^-p$ (right) differential cross section at $|t| = 0.1$~GeV$^2$ as a function of $\sqrt{s}$.
The solid, dashed and dotted curves represent the ratios for the Pomeron exchange, the Reggeon exchange and the cross term, respectively.
}
\label{contribution_pip_dcs}
\end{figure}

Finally, we show the energy dependence of the Pomeron and Reggeon exchange contribution.
We numerically evaluate the both contributions to the total cross section and divide by the total magnitude separately.
We define these as the contribution ratios, $R^{\pi^+p}_{\rm{tot}}$ and $R^{\pi^-p}_{\rm{tot}}$, for the $\pi^+ p$ and $\pi^- p$ scattering, respectively.
The $\sqrt{s}$ dependence of these ratios is displayed in Fig.~\ref{contribution_pip_tcs}.
It is seen that the ratio for the Pomeron exchange contribution increases with $\sqrt{s}$, and it is opposite for the Reegeon exchange contribution.
This behavior is common to both the $\pi^+p$ and $\pi^-p$ cases, but the Reggeon contribution in the $\pi^-p$ case is slightly larger than that in the $\pi^+p$ case.
For the both cases, the Reggeon contribution almost completely vanishes around $\sqrt{s} \sim 100$~GeV.
Then we perform the similar analysis for the differential cross sections, defining the contribution ratios, $R^{\pi^+ p}_{\rm{diff}}$ and $R^{\pi^- p}_{\rm{diff}}$, and display the results in Fig.~\ref{contribution_pip_dcs}.
Differently from the total cross section case, there is a contribution from the cross term.
The qualitative behavior of the ratios for the Pomeron and Reggeon contribution is similar to the total cross section case, and the ratio for the cross term decreases as $\sqrt{s}$ increases.
Not only the ratio for the Reggeon contribution but also that for the cross term in the $\pi^-p$ case are slightly larger than those in the $\pi^+p$ case.
From these results it is found that considering the Reggeon exchange contribution is necessary to correctly describe the cross sections, unless the energy is high enough.

%%%%%%%%%%%%%%%%%%%%%%%%%%%%%%%%%%%%%%%%%%%%%%%%%%%%%%%%%%%%%%%%%%%%%%%%%%%%%%%
\section{Conclusion}
\label{sec:conclusion}
%%%%%%%%%%%%%%%%%%%%%%%%%%%%%%%%%%%%%%%%%%%%%%%%%%%%%%%%%%%%%%%%%%%%%%%%%%%%%%%
We have studied the elastic $\pi p$ and $\pi\pi$ scattering in a holographic QCD model, focusing on the Regge regime.
To obtain the total and differential cross sections, we have taken into account the Pomeron and Reggeon exchange, which are described by the Reggeized $2^{++}$ glueball and vector meson propagator, respectively.
For the Pomeron-hadron couplings, the gravitational form factors, which can be calculated with the bottom-up AdS/QCD models, are utilized.
Our model setup involves nine parameters in total, but for six of them the values obtained in the preceding works can be employed, by virtue of the universality of the Pomeron and Reggeon.

We have determined the three adjustable parameters with the experimental data of the $\pi p$ total cross sections, and shown that those data can be well described within the model.
Once those parameters are determined, the $\pi \pi$ total cross sections and the $\pi p$ and $\pi \pi$ differential cross sections can be predicted without any additional parameters.
We have presented our predictions for all the charged pion combinations for both the total and differential cross sections.
Then we have demonstrated the comparisons between our predictions and the experimental data for the $\pi p$ differential cross sections, and shown that our calculations are consistent with the data.
Furthermore, we have investigated the energy dependence of the Pomeron and Reggeon contribution, focusing on the contribution ratios.
Our results indicate that taking into account the Reggeon exchange contribution is necessary to correctly describe the cross sections, unless the energy is high enough.

In this study we have extended the previous work~\cite{Liu:2022out}, in which only the Pomeron contribution was considered, and found that in a wider $\sqrt{s}$ range the present model can well describe both the total and differential cross sections.
The results presented in this paper show the predictive ability of the model, which may be useful for analyzing other high energy forward scattering processes.
However, it is also true that the experimental data used for the comparisons concentrate in some narrow kinematic region.
In particular, the data in the high energy region are quite scarce.
It is expected that the future data will help to further test the model and to deepen our understanding of the nonperturbative nature of the underlying strong interaction.

%%%%%%%%%%%%%%%%%%%%%%%%%%%%%%%%%%%%%%%%%%%%%%%%%%%%%%%%%%%%%%%%%%%%%%%%%%%%%%%
\section*{Acknowledgments}
%%%%%%%%%%%%%%%%%%%%%%%%%%%%%%%%%%%%%%%%%%%%%%%%%%%%%%%%%%%%%%%%%%%%%%%%%%%%%%%
The work of A.W. was supported by the start-up funding from China Three Gorges University.
A.W. is also grateful for the support from the Chutian Scholar Program of Hubei Province.

\bibliography{references}

\end{document}